\documentclass{article}
\usepackage{kr}
\usepackage{times}
\usepackage{soul}
\usepackage{url}
\usepackage[hidelinks]{hyperref}
\usepackage[utf8]{inputenc}
\usepackage[small]{caption}
\usepackage{graphicx}
\usepackage{amsmath}
\usepackage{amsthm}
\usepackage{booktabs}
\usepackage{algorithm}
\usepackage{amsfonts}
\usepackage{bm}
\usepackage{dirtytalk}
\usepackage{cleveref}
\usepackage{comment}

\usepackage{mymacros}
\usepackage{tikz}
\usetikzlibrary{positioning,decorations.pathreplacing,shapes}
\usepackage{tikz-cd}
\usepackage{algpseudocode}

\usepackage{mathtools}

\def\bx{\bm{x}}
\def\fomc{\mbox{\sc fomc}}
\def\wfomc{\mbox{\sc wfomc}}
\def\w{\mbox{\sc w}}

\def\bk{{\bm{k}}}
\def\bh{\bm{h}}

\def\bt{{\bm{t}}}
\def\bv{{\bm{v}}}
\def\bp{{\bm{p}}}
\def\bs{{\bm{s}}}

\newtheorem{definition}{Definition}
\newtheorem{remark}{Remark}
\newtheorem{proposition}{Proposition}

\newtheorem{example}{Example}
\newtheorem{theorem}{Theorem}

\title{Towards Counting Markov Equivalence Classes with Logical Constraints}

\author{%
Davide Bizzaro$^1$\and
Luciano Serafini$^{1}$\And 
Sagar Malhotra$^2$
\affiliations
$^1$Fondazione Bruno Kessler\\
$^2$TU Wien\\
\emails
dbizzaro@fbk.eu, sagar.malhotra@tuwien.ac.at, serafini@fbk.eu
}

\begin{document}

\maketitle

\begin{abstract}

We initiate the study of counting Markov Equivalence Classes (MEC) under logical constraints. MECs are equivalence classes of Directed Acyclic Graphs (DAGs) that encode the same conditional independence structure among the random variables of a DAG model. Observational data can only allow to infer a DAG model up to Markov Equivalence. However, Markov equivalent DAGs can represent different causal structures, potentially super-exponentially many. Hence, understanding MECs combinatorially is critical to understanding the complexity of causal inference. In this paper, we focus on analysing MECs of size one, with logical constraints on the graph topology. We provide a polynomial-time algorithm (w.r.t.\ the number of nodes) for enumerating essential DAGs (the only members of an MEC of size one) with arbitrary logical constraints expressed in first-order logic with two variables and counting quantifiers (C$^2$). Our work brings together recent developments in tractable first-order model counting and combinatorics of MECs.

\end{abstract}
\section{Introduction}
We provide a polynomial time algorithm for counting a sub-class of Markov Equivalence Classes (MECs) of Directed Acyclic Graphs (DAGs) with bounded indegree. MECs are equivalence classes of DAGs that represent the same conditional independence structure amongst the random variables of a DAG model, such as a Bayesian Network. However, the causal structure of a DAG model is uniquely defined by a single DAG. Hence, knowing the sizes (i.e., the number of DAGs in the MECs) and the number of MECs is fundamental to  understanding the complexity of causal inference. We are primarily concerned with counting the number of MECs. Although this problem has been extensively investigated in the literature \cite{Gillispie2001,GILLISPIE2006}, the problem of adding logic-based background knowledge, while counting MECs, has not been investigated formally. In this paper we make first steps in this direction by providing an algorithm that computes the number of MECs of size one under arbitrary logical constraints expressed in first-order logic with two variables and counting quantifiers (C$^2$). 

MECs of size one are completely characterized by \emph{essential DAGs} --- the only members of the class. A key feature of essential DAGs is that they can be uniquely characterized in terms of the DAG topology \cite{andersson_characterization_1997}. This insight has been used to extend results for counting DAGs to also enumerate essential DAGs \cite{enumeration_essential_DAGs}. The key focus of this paper is to expand these results to the case with logical constraints. For instance, we may be interested in counting essential DAGs with colored nodes (potentially representing properties like gender or occupation) while also necessitating, for example, that each node has at least one parent of each color. Such constraints can be easily expressed in first-order logic. Hence, our problem reduces to counting models of a first-order logic formula where one of the predicate symbols is axiomatized to represent an essential DAG.  

Motivated by applications in statistical relational AI \cite{SRL_LISA,SRL_LUC}, there has been significant interest in First-Order Model Counting (FOMC), that is counting the models of a First-Order Logic (FOL) sentence \cite{Symmetric_Weighted}. These results have identified rich logical fragments in which FOMC can be computed in polynomial time w.r.t the domain cardinality. One of the most important such fragment is FOL with two variables and counting quantifiers \cite{kuzelka2020,AAAI_Sagar}. Recent results \cite{malhotra2023lifted,Tree} have expended this fragment with graph theoretic constraints that axiomatize a given predicate to be a tree, connected graph, or a DAG. In this paper we will combine the techniques developed for counting essential DAGs with those developed for FOMC. 
As a result, we will show that the FOMC of any C$^2$ formula where one of the predicates is axiomatized to be an essential DAG with bounded indegrees can be done in polynomial time w.r.t the number of elements in the logical domain.

We first begin with discussing related works in Section 2, followed by background on MECs, essential DAGs, first-order logic, and FOMC in Section 3. We introduce our main results on counting essential DAGs with arbitrary C$^2$ constraints in Section 4. In Section 5 we provide some initial experiments, and compute the number of essential DAGs by their indegrees. Finally, we conclude in Section 6.



\section{Related Work}
\label{sec: rel_works}
\cite{DBLP:journals/ml/HeckermanGC95} recognized that Bayesian Networks can only be learnt up to Markov equivalence from observational data. \cite{andersson_characterization_1997} provide a characterization of MECs based on \emph{skeleton} (cf. Defintion~\ref{def:skeleton}) and \emph{immoralities} of a DAG (cf. Definition~\ref{def:immoralities}). Given such a topological characterization, a number of results analyzing size and number of MECs has been developed \cite{Gillispie2001,GILLISPIE2006,steinsky2004asymptotic,ghassami2019counting,talvitie2019counting}. \cite{Gillispie2001} empirically show that MECs of size one are the most common MECs. Hence, the number MECs of size one, i.e., the number of essential DAGs is a good indicator for lower bound on the total number of MECs. \cite{enumeration_essential_DAGs} 
provides an algorithm for enumerating essential DAGs by the number of nodes --- which forms the basis of the results presented in this paper. 

For FOMC there have been a series of results \cite{First_Order_Prob_Inf,de_Salvo,Symmetric_Weighted,kuusisto2018weighted,kuzelka2020,DilkasB23,AAAI_Sagar,Symmetric_Weighted,Jaeger_lifted,Tree,Linear_Order_Axiom} that have identified a rather rich fragment of first-order logic to admit tractable FOMC. In particular, our work draws most from the recent techniques of \emph{counting by splitting} that expand tractable FOMC fragments with graph theoretic constrains \cite{malhotra2023lifted}. 











\section{Background}\label{sec: background} 



\subsection{Notation and Basic Combinatorics} 
Let $[n]$ represent the set of integers $\{1,\dots,n\}$. In situations where the set of integers $[n]$ is clear from the context, we will utilize $[\overline{m}]$ to denote the set $[n] \backslash [m]=\{m+1, \dots, n\}$. Non-negative integer vectors are denoted by bold font letters, such as $\mathbf{k}$, while the corresponding regular font letters with an additional index, such as $k_i$, represent the components of the vectors. As an example, $\mathbf{k} = \langle k_1,...,k_u \rangle$ represents a vector of $u$ non-negative integers. The notation $|\mathbf{k}|$ indicates the sum of all the components of $\mathbf{k}$. Hence, a summation of the form $\sum_{|\bm{k}|=n}f(\bm{k})$, for any function $f$, runs over all possible $\bm{k}$ such that $\sum_{i\in[u]}k_i = n$. Vectors addition and subtraction are performed element-wise.

\textit{Multinomial coefficients} are denoted as follows:
$$\binom{|\bk|}{\bk}= \binom{|\bk|}{k_1,...,k_u} \coloneqq  \frac{|\bk|!}{\prod_{i\in [u]} k_i !}$$
The one above counts the number of ways to distribute $|\bk|$ objects into $u$ distinct categories, in such a way that each category $i$ contains exactly $k_i$ objects. 
It can be computed in polynomial time with respect to $|\bk|$. Moreover, the number of non-negative integer vectors $\bk$ of fixed length $u$ such that $|\bk| = n$ is bounded by a polynomial in $n$, that is $n^u$.

Given a set of finite sets $\{A_i\}_{i\in [n]}$, let $A_{J}:= \bigcap_{j\in J}A_j$ for any subset $J\subseteq [n]$. The \emph{principle of inclusion-exclusion (PIE)} states that:
\begin{equation}
    \label{P_IE}
    \Big|\bigcup_{i}A_i\Big| = \sum_{\emptyset \neq J \subseteq [n]} (-1)^{|J|+1} \big| A_{J} \big|
\end{equation}
If $A_{J}$ and $A_{J'}$ have the same cardinality $|A_{[m]}|$ whenever $J,J' \subseteq [n]$ are such that $|J| = |J'| = m$, then: 
\begin{equation}
    \label{P_IE_Symm}
    \Big|\bigcup_{i}A_i\Big| = \sum_{m=1}^{n} (-1)^{m+1} \binom{n}{m} \big |A_{[m]} \big |
\end{equation}


\subsection{Markov Equivalence and Essential Graphs}
Two DAGs are said to be Markov equivalent if they determine the same statistical model, i.e.\ the same conditional independences among the variables of a Bayesian network. Importantly, Markov equivalence admits a well-known structural characterization, given by \cite[Theorem 1]{Verma1990EquivalenceAS} and reported below, after the two definitions. 
\begin{definition}[Skeleton]
\label{def:skeleton}
    The \emph{skeleton} of a DAG is the underlying undirected graph, i.e.\ the graph obtained replacing the directed edges with undirected ones. 
\end{definition}
\begin{definition}[Immoralities]
\label{def:immoralities}
    The \emph{immoralities} of a DAG are the triples of vertices $(a,b,c)$ such that the subgraph on $\{a,b,c\}$ is $a\rightarrow b \leftarrow c$ (so that the parents $a$ and $b$ of $c$ are ``unmarried'').
\end{definition}
\begin{theorem}[\cite{Verma1990EquivalenceAS}]
\label{th:characterization_MECs}
    Two DAGs are Markov equivalent if and only if they have the same skeleton and the same immoralities.
\end{theorem}

\begin{example}[\cite{andersson_characterization_1997}]\label{ex:MECs}
The following are the DAGs with ``a square skeleton'' and an immorality $(a,b,c)$: 
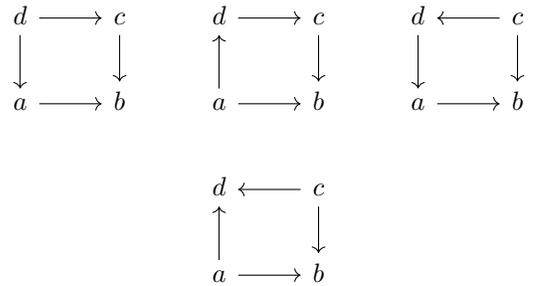
\begin{equation*}
\begin{tikzcd}
d \arrow[d] \arrow[r] & c \arrow[d] & d \arrow[r]           & c \arrow[d]           & d \arrow[d] & c \arrow[l] \arrow[d] \\
a \arrow[r]           & b           & a \arrow[u] \arrow[r] & b                     & a \arrow[r] & b                     \\
                      &             & d                     & c \arrow[d] \arrow[l] &             &                       \\
                      &             & a \arrow[r] \arrow[u] & b                     &             &                      
\end{tikzcd}
\end{equation*}
The three DAGs in the first line have no other immorality apart for $(a,b,c)$, so they constitute a MEC of size $3$. The DAG on the second line has another immorality $(a,d,c)$, so it constitute a MEC of size $1$ (i.e., an essential DAG).
\end{example}

While useful, \cref{th:characterization_MECs} does not provide a practical way to compute the MEC of a DAG, or to count MECs. This can be better done by characterizing essential graphs, which were introduced for this purpose in \cite{andersson_characterization_1997} and are graphs (admitting both directed and undirected edges) that uniquely represent a MEC. 

\begin{definition}[Essential Graph]
    The essential graph $D^*$ of a DAG $D$ is the smallest graph containing all the DAGs in the same MEC of $D$. In other words, $D^*$ is the graph with the same skeleton as $D$, but where an edge is directed if and only if it has the same orientation in all the DAGs equivalent to $D$.
\end{definition}

\begin{definition}[Essential DAG]
    An essential DAG is a DAG that coincides with its essential graph, i.e., a DAG that is the unique element of its MEC.
\end{definition}

\begin{example}    The following are the essential graphs of the two MECs in \cref{ex:MECs}:
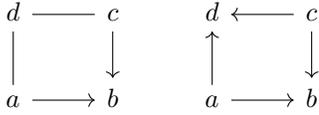
    \begin{equation*}
\begin{tikzcd}
d \arrow[d, no head] \arrow[r, no head] & c \arrow[d] & d                     & c \arrow[l] \arrow[d] \\
a \arrow[r]                             & b           & a \arrow[u] \arrow[r] & b                    
\end{tikzcd}
    \end{equation*}
The last one is a DAG, so it is an essential DAG.
\end{example}

A first structural characterization of essential graphs was provided by  \cite[theorem 4.1]{andersson_characterization_1997}, and used for counting in \cite{Gillispie2001}.
Here we present a method for counting essential DAGs, so we use their structural characterization, as given in \cite[corollary 4.2]{andersson_characterization_1997} and used for counting in \cite{enumeration_essential_DAGs}.

\begin{definition}[Protected Edge]
    In a DAG, an edge $a \to b$ is \emph{protected} if the the set of parents of $a$ is different from the set of parents of $b$ when $a$ is excluded. In other words, $a \to b$ is \emph{not} protected if every parent of $a$ is a parent of $b$ and $b$ has no additional parents apart for $a$.
\end{definition}

\begin{example}
    The edges starting from $d$ in the first two graphs of \cref{ex:MECs} and the one arriving to $d$ in the third are not protected. On the other hand, the edges in the graph in the second line are all protected.
\end{example}

\begin{proposition}[\cite{andersson_characterization_1997}]\label{prop:characterization_essential_DAGs}
    A DAG is essential if and only if all its edges are protected.
\end{proposition}

\subsection{Counting Essential DAGs}
Following \cite{enumeration_essential_DAGs}, we present the recursive formula for counting essential DAGs. It is similar to the one for counting DAGs in \cite{robinson1973counting}, but uses sinks (i.e.\ nodes with outdegree $0$) instead of sources (i.e.\ nodes with indegree $0$), as sinks have the following useful property: 
\begin{remark}\label{observation2_essential}
The DAG obtained removing a sink from an essential DAG is still essential (since all the edges remain protected). 
\end{remark}
Hence, we assume the set of nodes to be $[n]$, and for each $i\in [n]$, we let $A_i$ be the set of essential DAGs for which the node $i$ is a sink. Moreover, let $a_n$ denote the number of essential DAGs on $n$ labeled nodes. 
Since every DAG has at least a sink, $a_n$ is equal to $|\bigcup_{i\in [n]}A_i|$, and the principle of inclusion-exclusion tells us that 
\begin{equation}
    \label{eq:essential_inclusion_exclusion}
    a_n = \Big|\bigcup_{i\in [n]} A_i \Big| = \sum_{m=1}^{n} (-1)^{m+1} \binom{n}{m} |A_{[m]}|
\end{equation}
where $A_{[m]}\coloneqq \cap_{i\in [m]} A_i$.
In order to count $|A_{[m]}|$, we observe the following:

\begin{remark}\label{observation3_essential}
If $\omega\in A_{[m]}$, then there are no edges in $\omega$ between the nodes in $[m]$. Therefore, in order to extend an essential DAG on $[\overline m]$ to an essential DAG in $A_{[m]}$, we can only draw outgoing edges from $[\overline m]$ to $[m]$. Moreover, the outgoing edges that we can draw must be protected. It means that the set of parents of a node $i\in [m]$ can be any subset of $[\overline m]$ that is not made of a node plus exactly its set of parents. Therefore, from the $2^{n-m}$ possible subsets of $[\overline m]$, we need to exclude exactly one subset for each  node in $[\overline m]$, giving $2^{n-m} - (n-m)$ choices for each node $i$ in $[m]$. Thus, any essential DAG on $[\overline m]$ can be extended to $(2^{n-m} - (n-m))^m$ different essential DAGs in $A_{[m]}$.
\end{remark}

Since the number of possible essential DAGs on $[m]$ is by definition $a_{n-m}$, \cref{observation3_essential} tells us that there are $(2^{n-m} - (n-m))^m a_{n-m}$ essential DAGs in $A_{[m]}$ that are obtained extending DAGs on $[\overline m]$. But these are all the essential DAGs in  $A_{[m]}$, because if $\omega\in A_{[m]}$, then the subgraph of $\omega$ restricted to $[\overline m]$ is an essential DAG (\cref{observation2_essential}). We conclude that $|A_{[m]}| = (2^{n-m} - (n-m))^m a_{n-m}$. Substituting this value into \cref{eq:essential_inclusion_exclusion} gives the recursive formula we wanted:
\begin{equation}
    \label{eq:counting_essential_DAGs}
    a_n = \sum_{m=1}^{n} (-1)^{m+1} \binom{n}{m} (2^{n-m} - (n-m))^m a_{n-m}
\end{equation}
where $a_0 = 1$. 
Notice that, using this recursion with dynamic programming, one can easily derive an algorithm with polynomial complexity for computing the number of essential DAGs on $n$ nodes. 

\subsection{First-Order Logic} 

We use a first-order logic (FOL) language $\mathcal{L}$ that does not include function symbols and consists of a set of variables $\mathcal{V}$ and a set of relational symbols $\mathcal{R}$. If a relational symbol $R$ has arity $k$, we denote it as $R/k$. Given a tuple $(x_1, \ldots, x_k) \in \mathcal{V}^{k}$ and a relational symbol $R/k \in \mathcal{R}$, the expression $R(x_1, \ldots, x_k)$ is called an \emph{atom}. A \emph{literal} is either an atom or the negation of an atom. A \emph{formula} is created by connecting atoms using Boolean operators (such as $\neg$, $\lor$, and $\land$) and quantifiers ($\exists x_i.$ for existential quantification and $\forall x_i.$ for universal quantification), following the syntax rules of FOL. The \emph{free} variables in a formula are those not bound by any quantifier. We use $\phi(x_1, \ldots, x_k)$ to represent a formula with free variables ${x_1, \ldots, x_k}$; while a formula without any free variables is called a \emph{sentence} and is usually denoted with capital Greek letters. 
Sentences in $\mathcal{L}$ are interpreted over a domain $\Delta$, which is a set of constants. Ground atoms (and ground literals) are formed by replacing all variables in the atoms (or literals) with constants from the domain. Notice that for a predicate $R/k$ and a domain $\Delta$ of size $n$, there are $n^k$ possible ground atoms of the form $R(a_1, \ldots, a_k)$, where $(a_1, \ldots a_k) \in \Delta^k$.
An \emph{interpretation} $\omega$ over a finite \emph{domain} $\Delta$ is a truth assignment for all ground atoms. Assuming Herbrand semantics \cite{Herbrand_Logic}, $\omega$ is considered an interpretation or model of $\Psi$ if $\omega \models \Psi$. 


Let $\Delta'$ be a subset of a domain $\Delta$, and let $\omega$ be an interpretation. We denote by $\omega \downarrow \Delta'$ the interpretation on $\Delta'$ induced by $\omega$. It means that $\omega \downarrow \Delta'$ is an interpretation for the language with domain $\Delta'$, and it includes all and only the ground atoms included in $\omega$ and expressing only domain elements in $\Delta'$. 
Furthermore, we use $\omega_{R}$ to denote the partial interpretation of $\omega$ when restricted to the relation $R$. If $R$ is a unary relation, then $\omega_{R}$ can be viewed as a subset of the domain. If instead $R$ is a binary relation, then $\omega_{R}$ can be viewed as a directed graph, where an edge between the domain elements $c$ and $d$ exists if and only if $\omega \models R(c,d)$. This is what allows us to say that a binary relation forms an essential DAG.

\begin{example}
    \label{ex: projection}
    Consider a language with domain $\Delta = [4]$ and only three relational symbols, namely $R/2$, $B/2$, and $G/1$. To represent an interpretation $\omega$, we utilize a multigraph with colored nodes and arrows: we draw a red (resp.\ blue) directed edge from a node $c$ to a node $d$ whenever $\omega \models R(c,d)$ (resp.\ $\omega \models B(c,d)$); and we color a node $c$ with green whenever $\omega \models G(c)$. Let's examine the following interpretation $\omega$:
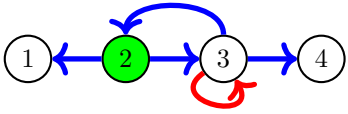
\medskip
\begin{center}
\begin{tikzpicture}[node distance={13mm}, thick, main/.style = {draw, circle}] 
        \node[main, fill = green] (1) {$2$}; 
        \node[main] (2) [left of=1] {$1$}; 
        \node[main] (3) [right of=1] {$3$}; 
        \node[main] (4) [right of=3] {$4$};  
        \draw[blue, ->,line width = 2pt] (1) -- (2); 
        \draw[blue,->,line width = 2pt] (1) to (3);
        \draw[blue,->,line width = 2pt] (3) to [out=90,in=90,looseness=1] (1);
        \draw[red,->,line width = 2pt] (3) to [out=215,in=300,looseness=4] (3); 
        \draw[blue,->,line width = 2pt] (3) -- (4); 
        \end{tikzpicture}
    \end{center}
\noindent
For such an interpretation, $\omega'\coloneqq \omega \downarrow [2]$ and $\omega''\coloneqq \omega \downarrow [\overline{2}]$ are given respectively as
\medskip
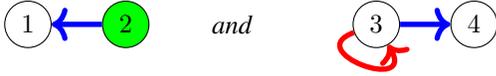
\begin{center}
    \begin{tikzpicture}[node distance={13mm}, main/.style = {draw, circle}] 
        \node[main, fill=green] (1) {$2$}; 
        \node[main] (2) [left of=1] {$1$}; 
        \node[main] (3) [right = 4cm of 2] {$3$}; 
        \node[main] (4) [right of=3] {$4$};  
        \draw[blue,->,line width = 2pt] (1) -- (2);
        \node[fill=none,align=center]{\hskip 8em and};
        \draw[red,->,line width = 2pt] (3) to [out=200,in=300,looseness=4] (3); 
        \draw[blue,->,line width = 2pt] (3) -- (4); 
    \end{tikzpicture}
\end{center}
\end{example}

Counting by splitting relies on decomposing the domain into two disjoint parts, and then makes use of partial interpretations for each part. So, let $\Delta'$ and $\Delta''$ be two disjoint sets of domain constants, and let us denote their union by $\Delta' \uplus \Delta''$. Given an interpretation $\omega'$ on $\Delta'$ and an interpretation $\omega''$ on $\Delta''$, the partial interpretation $\omega' \uplus \omega''$ on $\Delta' \uplus \Delta''$ is constructed in this way: it assigns truth values to ground atoms over $\Delta'$ according to $\omega'$, and to ground atoms over $\Delta''$ according to $\omega''$. However, the ground atoms involving domain constants from both $\Delta'$ and $\Delta''$ remain uninterpreted. This is clarified by the following example.



\begin{example}
    \label{ex: extension}
    Let us have the same language, representation and the interpretation $\omega$ as in \cref{ex: projection}. Now let us create two other interpretations $\omega'$ and $\omega''$, such that 
    $\omega' = \omega \downarrow [2]$ and $\omega'' = \omega \downarrow [\bar{2}]$ respectively on the domains $[2]$ and $[\overline{2}]$: 
The partial interpretation $\omega' \uplus \omega''$ is created as follows:
\medskip
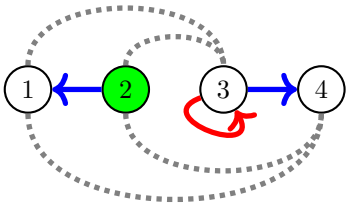
\begin{center}
\begin{tikzpicture}[node distance={13mm}, thick, main/.style = {draw, circle}] 
        \node[main, fill = green] (1) {$2$}; 
        \node[main] (2) [left of=1] {$1$}; 
        \node[main] (3) [right of=1] {$3$}; 
        \node[main] (4) [right of=3] {$4$};  
        
        \draw[blue, ->,line width = 2pt] (1) -- (2); 
        \draw[red,->,line width = 2pt] (3) to [out=200,in=300,looseness=4] (3); 
        \draw[blue,->,line width = 2pt] (3) -- (4); 
        
        \draw[gray, dotted, line width = 2pt] (1) to [out=90,in=90,looseness=1] (3);
        \draw[gray, dotted, line width = 2pt] (2) to [out=90,in=90,looseness=1] (3);
        \draw[gray, dotted, line width = 2pt] (1) to [out=270,in=270,looseness=1] (4);
        \draw[gray, dotted, line width = 2pt] (2) to [out=270,in=270,looseness=1] (4);
        \end{tikzpicture}
    \end{center}
The dotted lines 
represent the fact that the ground atoms concerning those pairs of nodes ($R(1,3)$, $R(3,1)$, $B(1,4)$, etc.) are not interpreted in $\omega' \uplus \omega''$. 
A possible extension to a complete interpretation is $\omega$ in \cref{ex: projection}. 
In total, we can see that  $\omega' \uplus \omega''$ can be extended in $2^{16}$ ways, as we have two mutually exclusive choices for assigning truth values to each of the $16$ uninterpreted ground atoms.  
\end{example}

\subsubsection{FO$^2$ and its extensions} 
Notice that the condition for a DAG to be essential formulated in \cref{prop:characterization_essential_DAGs} is expressed by the following FOL sentence:
\begin{equation*}
    \forall xy. Rxy \to \exists z. (Rzx \land \neg Rzy) \lor (Rzy \land \neg Rzx \land z\neq x)
\end{equation*}
where $R/2$ is the relational symbol representing the DAG.
However, the sentence has three variables, and (W)FOMC is known to be intractable ($\#$P-complete) in the general case of FOL formulas with three variables \cite{Symmetric_Weighted,Jaeger_lifted}. 
Given this limitation, extensive research has explored various fragments of FOL that allow for polynomial-time FOMC algorithms. The most basic one is FO$^2$, which restricts the set of variables to only two elements, typically denoted as $x$ and $y$.
One of the most useful and significant extensions of FO$^2$ is the one with \emph{counting quantifiers}, commonly denoted by C$^2$ \cite{COUNTING_REF}. The counting quantifiers are $\exists^{=k}$ (there exist exactly $k$), $\exists^{\geq k}$ (there exist at least $k$), and $\exists^{\leq k}$ (there exist at most $k$). The semantics of the first is defined by saying that $\omega \models \exists^{=k} x. \phi(x)$ for any $\phi$ and $\omega$ such that $\omega \models \phi(c)$ for exactly $k$ elements $c\in \Delta$; analogously for the others. 
Another important extension of FO$^2$ is the one with \emph{cardinality constraints}, which are constraints on the number of times the grounded predicates are interpreted to be true. For example, the cardinality constraint $|P| \geq 3$ impose the predicate $P$ to have at least three ground atoms interpreted as true (similarly to the C$^2$ formula $\exists^{\geq 3} x. P(x)$).

\subsubsection{Types, Tables and Consistency}
Our approach to WFOMC makes use of the notions of $1$-types, $2$-tables and $2$-types, as done in \cite{kuusisto2018weighted,ECML_PROJ}. 
Informally, in a given FOL language, \emph{$1$-types} represent the most specific (and therefore mutually exclusive) unary properties that individual elements of the domain can have. Formally, a $1$-type is the conjunction of a maximally consistent set of literals containing only one variable. Here, maximally consistent means that there are not any two literals directly contradicting each other, and that the set cannot be extended without losing this property.  

\begin{example}\label{ex:1-types}
    With only a unary predicate $U$ and a binary predicate $R$, there are four $1$-types: 
    \begin{align*}
    U(x) &\land R(x,x) \\
    U(x)&\land \neg R(x,x) \\
    \neg U(x)&\land R(x,x) \\
    \neg U(x)&\land \neg R(x,x)
    \end{align*}
    In general, each $1$-type is constructed by choosing one out of two literals for each relational symbol. Therefore, the number of $1$-types, denoted by $u$, is $2$ to the power of the number of predicates.
\end{example}

Throughout, we assume an arbitrary order on the set of $1$-types. 
Thus, we can use $i(x)$ to denote the $i^{th}$ $1$-type, and say that a domain constant $c$ \emph{realizes} the $i^{th}$ $1$-type under the interpretation $\omega$ if $\omega \models i(c)$. Additionally, we use $u$ to denote the number of $1$-types in a language.

A \emph{$2$-table} is the conjunction of a maximally consistent set of literals containing exactly two variables, along with the condition $x \neq y$, if the two variables are named $x$ and $y$. 

\begin{example}\label{ex:2-tables}
    With only a unary predicate $U$ and a binary predicate $R$ (as in the previous example), there are four $2$-tables: 
    \begin{align*}
    R(x,y)\land R(y,x)&\land x \neq y \\
    R(x,y)\land \neg R(y,x) &\land x \neq y \\
    \neg R(x,y)\land R(y,x) &\land x \neq y  \\
    \neg R(x,y)\land \neg R(y,x) &\land x \neq y \\
    \end{align*}
    \noindent In general, each $2$-table is constructed by choosing one out of four possibilities for each binary relation. Therefore, the number of $2$-tables, denoted by $b$, is $4$ to the power of the number of binary relations.
\end{example}

Similarly to $1$-types, we assume an arbitrary order also for $2$-tables. Hence, we use $l(x,y)$ to denote the $l^{th}$ $2$-table, and state that an ordered pair of domain constants $(c,d)$ \emph{realizes} the $l^{th}$ $2$-table under the interpretation $\omega$ if ${\omega \models l(c,d)}$. The number of $2$-tables in a given language is denoted by $b$.

A \emph{$2$-type} is the conjunction of a maximally consistent set of literals containing one or two variables, along with the condition $x \neq y$. Equivalently, a $2$-type is a quantifier-free formula of the form $i(x)\land j(y) \land l(x,y)$, where $i(x)$ and $j(y)$ are $1$-types and $l(x,y)$ is a $2$-table. To represent such a $2$-type, we use the notation ${ijl(x,y)}$. So, we say that an ordered pair of domain constants $(c,d)$ \emph{realizes} the $2$-type ${ijl(x,y)}$ under the interpretation $\omega$ if ${\omega \models ijl(c,d)}$. Moreover, we immediately see that the number of $2$-types is $u^2b$.


Having defined $1$-types and $2$-types, we can introduce some related concepts and results, needed for FOMC.

\begin{definition}[$1$-Types Cardinality Vector] Since we considered an order on the $1$-types, we can identify them with the elements of $[u]$. Then, an interpretation $\omega$ is said to have the $1$-types cardinality vector $\bk = \langle k_1,\dots,k_u \rangle$ if there are exactly $k_i$ domain elements $c$ such that $\omega \models i(c)$, for each $1$-type $i$. To indicate that $\omega$ has the $1$-types cardinality vector $\bk$ we write $\omega \models \bk$.
\end{definition}


It should be noted that, in any interpretation $\omega$, every element in the domain realizes one and only one $1$-type. Consequently, if a $1$-types cardinality vector $\bk$ is given, then the domain cardinality must be equal to $|\bk|$. Moreover, let us be given a fixed pair of $1$-types $i$ and $j$ with $i \neq j$, along with a $1$-types cardinality vector $\bk$; then there are $k_ik_j$ pairs of domain constants $(c,d)$ such that $\omega \models i(c) \land j(d)$. Similarly, for a given $1$-type $i$ and $1$-types cardinality vector $\bk$, there exist $\binom{k_i}{2}$ unordered pairs of distinct domain constants $\{c,d\}$ that satisfy $\omega \models i(c) \land i(d)$.


\begin{definition}[$2$-Type Consistency]
    \label{def: 2-type_consistency}
    For any quantifier-free formula $\phi(x,y)$, we define 
    \begin{equation*}
    \phi(\{x,y\})\coloneqq \phi(x,x)\land \phi(x,y)\land \phi(y,x)\land \phi(y,y)\land x \neq y
    \end{equation*}
    With this notation, we can now express the notion of $2$-type consistency: we say that a $2$-type $ijl(x,y)$ is consistent with an FO$^2$ sentence $\forall xy. \phi(x,y)$, where $\phi(x,y)$ is quantifier-free, if
    \begin{equation}
      \label{2_type_consistency}
      ijl(x,y) \models \phi(\{x,y\})
    \end{equation}  
  The entailment is checked by assuming a propositional language whose atomic formulas are the atoms in the FOL language.
  \end{definition}

Notice that, in an interpretation $\omega \models \forall xy. \phi(x,y)$, a pair of domain constants $(c,d)$ can only realize the $2$-type $ijl(x,y)$ if the $2$-type is consistent with the formula $\forall xy. \phi(x,y)$. 
On the other hand, given an interpretation $\omega$, we must have that $\omega \models \forall xy. \phi(x,y)$, if the $2$-types realized by the pairs of domain constants are all consistent with $\forall xy. \phi(x,y)$. This is formalized in the following proposition, from \cite{malhotra2023lifted}.

\begin{proposition}[\cite{malhotra2023lifted}]
 \label{prop: ext}
    Let $\forall xy. \phi(x,y)$ be an FO$^2$ sentence where $\phi(x,y)$ is quantifier-free. Then, $\omega \models \forall xy. \phi(x,y)$ iff, for any pair of distinct domain constants $(c,d)$ such that $\omega \models ijl(c,d)$, we have that $ijl(x,y)$ is consistent with $\forall xy.\phi(x,y)$ (which is written $ijl(x,y) \models \phi(\{x,y\})$).
\end{proposition}

\subsection{Weighted First-Order Model Counting} 
\begin{definition}[Symmetric Weight Function]
    \label{def: symm}
    Let us be given an FOL language where $\mathcal{G}$ is the set of all ground atoms and $\mathcal{R}$ is the set of all relational symbols. A symmetric weight function~$(w, \overline w)$, with $w: \mathcal{R} \rightarrow \mathbb{R}$ and $\overline{w}: \mathcal{R} \rightarrow \mathbb{R}$, associates two real-valued weights  to each relational symbol. Then, the weight of an interpretation $\omega$ is defined as
    \begin{equation}
        \label{eq: symmweight}
        \w(\omega) \coloneqq \prod_{\substack{ \omega \models g \\ g \in \mathcal{G} \\}}w(pred(g)) \prod_{\substack{ \omega \models \neg g\\ g \in \mathcal{G} }}\overline{w}(pred(g))
    \end{equation} 
    where $pred(g)$ denotes the relational symbol in the ground atom $g$.
\end{definition}

From now on, whenever referring to weights, we intend symmetric weights. 
In order to make it easier to write the combinatorial formulas for WFOMC, we define a weight for each $1$-type $i(x)$, and one for each $2$-table $l(x,y)$: 
\begin{equation}\label{eq: w_i}
    w_i \coloneqq \prod_{\substack{i(x) \models  a \\ a \in \mathcal{A}}}w(pred(a)) \prod_{\substack{ i(x) \models \neg a \\ a \in \mathcal{A}}} \overline{w}(pred(a))
\end{equation}
and
\begin{equation}\label{eq: v_l}
    v_l \coloneqq  \prod_{\substack{l(x,y) \models  a \\ a \in \mathcal{A}}}w(pred(a)) \prod_{\substack{ l(x,y) \models \neg a \\ a \in \mathcal{A}}} \overline{w}(pred(a))
\end{equation}
where $\mathcal{A}$ denotes the set of atoms in the FOL language, $pred(a)$ denotes the relational symbol in the atom $a$, and $i(x) \models  a$ (resp.\ $l(x,y) \models  a$) indicates that $a$ is one of the conjuncts in $i(x)$ (resp. $l(x,y)$). 
Provided with these weights, we can forget the functions $w$ and $\overline w$ because all we need for WFOMC is in the weights of $1$-types and $2$-tables.

\subsubsection{WFOMC in \texorpdfstring{FO$^2$}{FO²}}
Let us define $\wfomc(\Psi,\bk)$ as the weighted model counting of the interpretations $\omega$ that satisfy the sentence $\Psi$ and have $1$-types cardinality vector $\bk$:
\begin{equation}
    \wfomc(\Psi,\bk) := \sum_{\omega \models \Psi \land \bk} \w(\omega)
\end{equation}
Remember that in this case the domain cardinality is $|\bk|$. So, if we want to compute the total WFOMC on $n$ nodes (denoted $\wfomc(\Psi,n)$), we can sum over all possible $1$-types cardinality vectors $\bk$ with $|\bk|=n$:
\begin{equation}
    \wfomc(\Psi,n) = \sum_{|\bk|=n}\wfomc(\Psi,\bk) 
\end{equation}

The approach that we use for polynomial-time WFOMC in FO$^2$ draws upon \cite{Symmetric_Weighted} and \cite{broeck2013}.
The first one shows how to perform WFOMC for sentences in FO$^2$ that are in \emph{Skolem normal form}, i.e.\ sentences of the form $\forall xy. \phi(x,y)$ where $\phi(x,y)$ is quantifier-free. The second provides a way to reduce the WFOMC of any FO$^2$ sentence to that of a sentence in Skolem normal form, for which the first method applies.

\begin{theorem}[\cite{Symmetric_Weighted}]
    \label{thm: beam}
For any FO$^2$ sentence $\forall xy. \phi(x,y)$ where $\phi(x,y)$ is quantifier-free, let us define the following quantity for any unordered pair of $1$-types $i,j$:
\begin{equation*} 
    r_{ij} \coloneqq \sum_{l\in[b]}n_{ijl}v_{l}
\end{equation*}
where the sum is over the $2$-tables, $v_l$ is defined in \cref{eq: v_l}, and
\begin{align*}
    n_{ijl} \coloneqq
    \begin{cases} 
        1 & \text{if \/ $ijl(x,y) \models \phi(\{x,y\})$} \\
        0 & \text{otherwise} \\
    \end{cases}
\end{align*} 
Then, for any $1$-types cardinality vector $\bk$,
   \begin{align*}
        \wfomc(\forall xy.\phi(x,y),\bk) =  \binom{|\bk|}{\bm{k}} \prod_{i\in [u]}w_i^{k_i}
        \prod_{\substack{{i\leq j \in[u]}}}\!\!\! r_{i j}^{\bk(i,j)} 
    \end{align*} 
where $w_i$ is defined in \cref{eq: w_i} and
\begin{align*}
  \bk(i,j) \coloneqq
  \begin{cases} 
      \frac{k_{i}(k_{i} - 1)}{2} & \text{if \/ $i=j$} \\
       k_{i}k_{j} & \text{otherwise} \\
     \end{cases}
\end{align*} 
\end{theorem}

Notice that the WFOMC formula in theorem can be computed in polynomial time w.r.t.\ the domain cardinality $|\bk|=n$. Since there are only polynomially many $\bk$ with $|\bk|=n$,  the WFOMC of $\forall xy.\phi(x,y)$ can be computed in polynomial time w.r.t.\ domain size $n$. 

In order to extend this approach to any FO$^2$ sentence, we can make use of \cite{broeck2013}, which shows that any FOL formula $\Phi$ with existential quantification can be modularly reduced to a WFOMC-preserving universally quantified FO$^2$ formula $\Phi'$, with additional new predicates and the use of negative weights. 
Modularity, as described in \cite{kuzelka2020}, means that 
\begin{equation*}
\wfomc(\Phi \land \Lambda,n) = \wfomc(\Phi' \land \Lambda,n)  
\end{equation*}  
for any constraint $\Lambda$, even not expressible in  FOL (such as e.g., the Tree axiom \cite{Tree}, the Linear Order axiom \cite{Linear_Order_Axiom} and the DAG axiom \cite{malhotra2023lifted}).

\subsubsection{WFOMC in \texorpdfstring{C$^2$}{C²}}
The problem of WFOMC for a C$^2$ sentence $\Phi$ can be solved by translating it into a problem in FO$^2$ \cite{kuzelka2020,AAAI_Sagar}. The central idea is the following. First, the WFOMC of $\Phi$ can be modularly transformed into the WFOMC of an FO$^2$ sentence $\Phi'$ with cardinality constraints $\Gamma$, within an extended vocabulary that includes additional predicates with weights of $1$ or $-1$. Then, $\wfomc(\Phi' \land \Gamma,\bk)$ can be computed in polynomial time relative to a $\wfomc(\Phi',\bk)$ oracle, using polynomial interpolation. For details, we refer to \cite{kuzelka2020}. It is important to notice that this reduction from a problem in C$^2$ to one in FO$^2$ is modular, and modularity allows it to be applied also when there are constraints inexpressible in first-order logic, like the one imposing a binary relation symbol to represent an essential DAG.

\section{Counting with Essential DAG Axiom}



We now derive the theory for WFOMC with the axiom $EssentialDAG(R, d)$, defined below and expressing the constraint that the relational symbol $R/2$ represent an essential DAG with indegrees bounded by $d$.

\begin{definition}
    \label{def:essential_DAG}
    Let $R/2$ be a relational symbol. We say that an interpretation $\omega$ is a model of $EssentialDAG(R, d)$, and write $\omega \models EssentialDAG(R, d)$, if $\omega_R$ forms an essential DAG and the indegrees of its nodes are less than or equal to $d$. 
\end{definition}

\begin{definition}
    \label{def: Psi_m_essential}
    Let $\Psi\coloneqq \Phi \land EssentialDAG(R, d)$, where $\Phi$ is an FOL sentence, and the domain is $[n]$. For any $m\in [n]$, we say that an interpretation $\omega$ is a model of $\Psi_{[m]}$, and write $\omega \models \Psi_{[m]}$, if $\omega$ is a model of $\Psi$ and the domain elements in $[m]$ have zero outdegree (i.e.\ are sinks) in the essential DAG formed by $\omega_R$.
\end{definition}

The following proposition gives us useful conditions on the restriction of a model of $\Psi_{[m]}$ to $[m]$ and to $[\overline m]$.

\begin{proposition}
    \label{prop:restriction_[m]_essential}
    Let $\Psi \coloneqq \forall xy. \phi(x,y) \land EssentialDAG(R, d)$ and $\Psi' \coloneqq \forall xy. \phi(x,y) \land \neg R(x,y)$, where $\phi(x,y)$ is quantifier-free. Let the domain be $[n]$, and let $1 \leq m \leq n$. If $\omega$ is a model of $\Psi_{[m]}$,  then $\omega \downarrow [m]  \models \Psi'$ and $\omega \downarrow [\overline{m}]  \models \Psi$. 
\end{proposition}
\begin{proof}
By definition of $\Psi_{[m]}$, since $\omega \models \Psi_{[m]}$, we have that $\omega \models \forall xy. \phi(x,y)$. Then, it must be that $\omega \downarrow [m] \models \forall xy. \phi(x,y)$ and $\omega \downarrow [\overline{m}] \models \forall xy. \phi(x,y)$. This fact --- that models of universally quantified formulas remain models when restricting the domain --- is a direct consequence  of what it means to be a model of a universally quantified formula in Herbrand semantics. 
Now, the fact that $\omega \models \Psi_{[m]}$ means that $\omega_{R}$ cannot have any edge in $[m]$. Hence, $\omega \downarrow [m] \models \Psi'$. Finally, because of \cref{observation2_essential}, we can conclude that if $\omega \models \Psi_{[m]}$, then $\omega \downarrow [\overline{m}] \models \Psi$.
\end{proof}


Following the approach of counting by splitting introduced in \cite{malhotra2023lifted}, we need to count the number of extensions of a model of $\Psi'$ on $[m]$ and a model of $\Psi$ on $[\overline m]$ to a model of $\Psi_{[m]}$ on the whole domain $[n]$.
The main problem is that the number of possible extensions when adding a sink depends on the configuration of edges in the essential DAG on $[\overline m]$. In deriving the formula for counting essential DAGs, for each sink in $[m]$ we excluded a subset of $[\overline m]$ for each node in $[\overline m]$ --- the subset given by the node and its parents. Now, the quantity to subtract depends on the graph made by $R$ on $[\overline m]$ and the different $1$-types of the nodes. 

To tackle this, we introduce new predicates $A_{(t_1,\dots,t_u)}/1$ saying \say{a node $y$ has $t_1$ $R$-parents of $1$-type $1$, $t_2$ $R$-parents of $1$-type $2$, ..., $t_u$ $R$-parents of $1$-type $u$}, i.e. satisfying the condition
\begin{align}
\begin{split}
    \label{eq:new_predicates_essential}
    \forall y. A_{(t_1,\dots,t_u)}(y)\leftrightarrow (\exists^{=t_1}x.1(x)\wedge R(x,y))\wedge\dots \\ \dots\wedge (\exists^{=t_u}x.u(x)\wedge R(x,y))
\end{split}
\end{align}
for any vector of natural numbers $\bt =(t_1,\dots,t_u)$ with $|\bt|\leq d$.
Let us denote by $T$ the set of all such vectors. Since the new predicates are all mutually exclusive, we can reduce the new ``extended'' $1$-types (i.e.\ the ones given the addition of the new predicates) to the pairs given by one of the $u$ original $1$-types and by a vector $\bt \in T$. In the following, 
the cardinality vectors $\bk, \bk', \bk''$ refer to these extended $1$-types in $[u]\times T$. In order to get the cardinality vectors for the original $u$ $1$-types, we introduce the function $\alpha\colon [n]^{[u]\times T}\to [n]^{[u]}$ such that
\begin{equation}
\alpha(\bk)_i = \sum_{t\in T} \bk_{(i,t)}
\end{equation}

Notice that adding these new predicates $A_{(t_1,\dots,t_u)}/1$ does not alter the FOMC, because  their truth values are completely determined by the truth values of the original predicates, given condition \eqref{eq:new_predicates_essential}. In formulas,
$$
\fomc(\Psi,n) = \sum_{|\bk|=n} \fomc(\Psi\wedge \eqref{eq:new_predicates_essential},\bk)
$$
for any $\Psi$. Moreover, with a slight abuse of notation, we will write $\fomc(\Psi\wedge \eqref{eq:new_predicates_essential},\bk)$ simply as $\fomc(\Psi,\bk)$ --- and similarly for $\Psi'$ and $\Psi_{[m]}$ ---, as if assuming that condition \eqref{eq:new_predicates_essential} is already encoded in the $EssentialDAG(R,d)$ axiom inside of $\Psi$ or in the fact that $\bk$ counts the extended $1$-types, not the original ones. The reason for this is that the satisfaction of \eqref{eq:new_predicates_essential} will be imposed by construction inside the induction step together with the essential DAG constrain, in a way that is different from the satisfaction of the FOL part of $\Psi$, $\Psi_{[m]}$ and $\Psi'$. 

The following proposition gives us a way to calculate the number of extensions to be used for computing $\fomc(\Psi_{[m]}, \bk)$. 

\begin{proposition}
    \label{prop: extensions_k_essential}
    Let $\Psi \coloneqq \forall xy. \phi(x,y) \land EssentialDAG(R, d)$ and $\Psi' \coloneqq \forall xy. \phi(x,y) \land \neg R(x,y)$, where $\phi(x,y)$ is quantifier-free, and let the domain be $[n]$. Since DAGs do not admit loops, we also assume without loss of generality that $\phi(\{x,y\}) \models \neg R(x,x)$. Let $\omega'$ be a model of $\Psi'$  on the domain $[m]$ and let $\omega''$ be a model of $\Psi$ on the domain $[\overline{m}]$. Moreover, let $k'_J$ (resp. $k''_I$) be the number of domain constants realizing the extended $1$-type $J$ (resp. $I$) in $\omega'$ (resp. $\omega''$).    
    Then, the number $N(\phi, \bk', \bk'')$ of extensions $\omega$ of $\omega' \uplus \omega''$  such that $\omega \models \Psi_{[m]}$ is
    \begin{equation}
        \label{eq:extensions_essential}
       \prod_{(j,\bt)\in[u]\times T} \left(
            g(\bk'', \bt) \prod_{i\in [u]} c_{ij}^{t_i} d_{ij}^{\alpha(\bk'')_i - \bt_i}
        \right)^{k'_{(j,\bt)}}
    \end{equation}
    where
    \begin{equation}
       g(\bk'', \bt) \coloneqq \prod_{i\in [u]} \binom{\alpha(\bk'')_i}{t_i} - \sum_{i\in [u]\mid \bt_i>0} \bk''_{(i,\bt^{-i})}
    \end{equation}
    and
    \begin{itemize}
        \item $\bt^{-i}\coloneqq(t_1,\dots, t_{i-1}, t_i - 1, t_{i+1}, \dots, t_u)$ is $\bt$ minus the vector with $1$ in position $i$ and $0$ in all other positions;
        \item $c_{ij}=\sum_{l\in [b]} c_{ijl}$
        \item $d_{ij}=\sum_{l\in [b]} d_{ijl}$
        \item $c_{ijl} = \begin{cases} 1 \text{ if } ijl(x,y)\models \phi(\{x,y\})\wedge R(x,y)\wedge \neg R(y,x) \\ 0  \text{ otherwise } \end{cases}$
        \item $d_{ijl} = \begin{cases} 1 \text{ if } ijl(x,y)\models \phi(\{x,y\})\wedge \neg R(x,y)\wedge \neg R(y,x) \\ 0  \text{ otherwise } \end{cases}$
    \end{itemize}  
\end{proposition}

\begin{proof}
    Notice that we can treat the sinks in $[m]$ independently, and for each one we have to count all the possible extensions of $\omega''$ to this node, while ensuring that the $R$-edges arriving to it are protected and that $\forall xy.\phi(x,y)$ remains satisfied. Let us fix one of the $\bk'_{(j,\bt)}$ sinks of extended $1$-type $(j,\bt)$. 
    Analogously to what was done with \cref{observation2_essential}, we first count all the ways in which we can choose the $R$-parents of the sink. This is given by $g(\bk'', \bt)$. Indeed, because the sink has extended $1$-type $(j,\bt)$, we have to choose $t_i$ $R$-parents among the $\alpha(\bk'')_i$ nodes in $[\overline m]$ that have original $1$-type $i$, for each $i\in [u]$. This gives us a total of $$\prod_{i\in [u]} \binom{\alpha(\bk'')_i}{t_i}$$ choices, from we need to exclude those with an unprotected edge. And the choices with an unprotected edge are those for which there is an $R$-parent of the sink which is the child of all the remaining $R$-parents of the sink (and of no other node). If such an $R$-parent has original $1$-type $i$, it must then have extended $1$-type $(i,\bt^{-i})$, because its parents are the parents of the sink minus itself. Vice versa, for every node of extended $1$-type $(i,\bt^{-i})$, for some $i$, the choice of it and its $R$-parents as the $R$-parents of the sink would have an unprotected edge. These choices are all distinct, so we need to exclude as many choices as the nodes in $[\overline m]$ which have extended $1$-type $(i,\bt^{-i})$, for some $i$, under $\omega''$. Thus, the number of choices that we exclude is given by $$\sum_{i\in [u]\mid \bt_i>0} \bk''_{(i,\bt^{-i})}$$
    This proves that $g(\bk'', \bt)$ counts the different ways in which we can choose the $R$-parents of a sink of extended $1$-type $(j,\bt)$, given $\omega''$.

    In order to count the models $\omega$ of $\Psi_{[m]}$ which extend $\omega' \uplus \omega''$, we consider the interpretations of the ground-atoms containing pairs $(c,d) \in [\overline m] \times [m]$, where $d$ has extended $1$-type $(j,\bt)$ under $\omega'$. Let us also fix the original $1$-type of $c$ as $i$. Because we want $\omega$ to be a model of $\Psi_{[m]}$, if $\omega\models ijl(c,d)$ for a two-table $l\in [b]$, then \cref{prop: ext} tells us that $ijl(c,d) \models \phi(\{c,d\})$.  Moreover, if $c$ is chosen as an $R$-parent of $d$, then $ijl(c,d)\models R(c,d) \wedge \neg R(d,c)$, otherwise $ijl(c,d)\models \neg R(c,d) \wedge \neg R(d,c)$. Therefore, $c_{ij}=\sum_{l\in [b]} c_{ijl}$ counts the interpretations of all the possible ground atoms containing the pair $(c,d)$ with $c$ being chosen as an $R$-parent of $d$. Similarly,  $d_{ij}=\sum_{l\in [b]} d_{ijl}$ counts the possible interpretations when $c$ is not being chosen as an $R$-parent of $d$. 

    For each $d\in [m]$ with extended $1$-type $(j,\bt)$, we have shown that we can chose its $R$-parents under $\omega$ in $g(\bk'',\bt)$ different ways. For each of these ways and for each original $1$-type $i$, we have $t_i$ pairs $(c,d)$ with $c$ being an $R$-parent of $d$ of $1$-type $i$, and $\alpha(\bk'')_i-t_i$ pairs $(c,d)$ with $c$ of $1$-type $i$ not being an $R$-parent of $d$. The former pairs admit $c_{ij}$ different interpretations, while the latter $d_{ij}$. Thus, in total, we have 
    $$g(\bk'', \bt) \prod_{i\in [u]} c_{ij}^{t_i} d_{ij}^{\alpha(\bk'')_i - \bt_i}$$ 
    possible extensions for each $d\in [m]$ with extended $1$-type $(j,\bt)$. There are exactly $\bk'_{(j,\bt)}$ such $d\in [m]$, and the extensions are independent of them, so we finally get the formula in \eqref{eq:extensions_essential}.
\end{proof}

The previous proposition was the tricky part; now the remaining is similar to that in \cite{malhotra2023lifted} for FOMC with DAG axiom. 

\begin{proposition}\label{prop:induction_essential}
    Let $\Psi \coloneqq \forall xy. \phi(x,y) \land EssentialDAG(R, d)$ and $\Psi' \coloneqq \forall xy. \phi(x,y) \land \neg R(x,y)$, where $\phi(x,y)$ is quantifier-free. Then:     
    \begin{align}
    \begin{split}
        \label{eq: k_k'_m_essential}
        &\fomc(\Psi_{[m]},\bk) = \\
        & =\sum_{\substack{\bk= \bk'+\bk''\\ |\bk'|=m}}\!\! N(\phi, \bk', \bk'') \fomc(\Psi',\bk')\fomc(\Psi,\bk'' )
    \end{split}
    \end{align}
    where $N(\phi, \bk', \bk'')$ is defined in \cref{prop: extensions_k_essential}.
\end{proposition}
\begin{proof}

    $\fomc(\Psi',\bk')$ is the FOMC of $\Psi'$ on $[m]$ with extended $1$-type cardinality vector $\bk'$. Similarly, $\fomc(\Psi,\bk'')$ is the FOMC of $\Psi$ on $[\overline{m}]$, with $1$-type cardinality vector $\bk''$. Due to \cref{prop: extensions_k_essential}, for any model $\omega'$ counted in $\fomc(\Psi',\bk')$ and any model $\omega''$ counted in $\fomc(\Psi,\bk'')$, the number of their extensions is given by $N(\phi, \bk', \bk'')$. Thus,
\begin{equation}
    \label{eq: wfomc_k'_k''}
     N(\phi, \bk', \bk'')\fomc(\Psi',\bk')\fomc(\Psi,\bk'' )
\end{equation}
gives us the FOMC of the models $\omega$ such that $\omega \models \Psi_{[m]}$ and $\omega \downarrow [m]  \models \Psi' \land \bk'$ and $\omega \downarrow [\overline{m}]  \models \Psi \land \bk''$. 
We want instead the FOMC of the models $\omega$ such that $\omega \models \Psi_{[m]} \land \bk$ and $\omega \downarrow [m]  \models \Psi'$ and $\omega \downarrow [\overline{m}]  \models \Psi$.
This is given by summing the quantity in \eqref{eq: wfomc_k'_k''} over all possible $1$-types cardinality vectors $\bk'$ and $\bk''$  that are consistent with $\bk$  (i.e.\ $\bk = \bk'+ \bk''$) and with the fact that $\bk'$ is interpreted in $[m]$ (so $|\bk'| = m$) and $\bk''$ is interpreted in $[\overline m]$ (so $|\bk''| = n-m$, already implied by $\bk = \bk'+ \bk''$, $|\bk'| = m$ and $|\bk|=n$). This is what is expressed in \eqref{eq: k_k'_m_essential}, and is justified by the fact that any $1$-types cardinality vector $\bk$ must decompose in such a way into two vectors $\bk'$ and $\bk''$, respectively on $[m]$ and $[\overline m]$.
\end{proof}

\begin{proposition}
    Let $\Psi \coloneqq \forall xy. \phi(x,y) \land EssentialDAG(R,d)$, where $\phi(x,y)$ is quantifier-free. Then:
    \begin{equation}
        \label{eq: DAG_Acyclic_essential}
        \fomc(\Psi ,\bk) = \sum_{m=1}^{|\bk|}(-1)^{m + 1}\binom{|\bk|}{m}\fomc(\Psi_{[m]},\bk)
    \end{equation} 
\end{proposition}
\begin{proof}
    Recall that we assume the domain be $[n]$, so $|\bk| = n$ and $m\in [n]$. For every $i\in [n]$, let $A_{i}$ be the set of models $\omega$ such that $\omega\models \Psi \land \bk$ and the domain element $i$ has zero outdegree in $\omega_R$. Since every DAG has at least one sink, $\bigcup_{i\in [n]}A_i$ contains all the models $\omega$ such that $\omega\models \Psi \land \bk$, and hence $\fomc(\Psi ,\bk) = |\bigcup_{i\in [n]}A_i)|$.  For any $J\subseteq [n]$, let $A_{J}\coloneqq  \bigcap_{j\in J}A_j$. Clearly we can use the principle of inclusion-exclusion as given in \cref{P_IE_Symm}, and get
    \begin{equation}
        \label{eq: PIE_Acyclic}
        \fomc(\Psi ,\bk)  = \sum_{m=1}^n (-1)^{m+1} \binom{n}{m}|A_{[m]}| 
    \end{equation}   
    $A_{[m]}$ is exactly the set of the models of $\Psi_{[m]}$, so $|A_{[m]}| = \fomc(\Psi_{[m]},\bk)$. 
Thus, equation \eqref{eq: PIE_Acyclic} becomes \eqref{eq: DAG_Acyclic_essential}.
\end{proof}



Combining \eqref{eq: DAG_Acyclic_essential} with \eqref{eq: k_k'_m_essential}, one can recursively compute $\fomc(\Psi ,\bk)$ in terms of  $\fomc(\Psi ,\bk'')$, with $\bk''$ smaller than $\bk$. This allow us to write algorithm \ref{alg:algorithm_FO_DAG} for computing the FOMC of $\Psi \coloneqq \forall xy. \phi(x,y) \land EssentialDAG(R, d)$. The algorithm initializes an array $A$ with $u\times |T|$ indices (i.e., as many as the extended $1$-types, and as many as the indices of $\bk$), which is a number bounded by $ud^u$. Then, it runs in lexicographical order over the vectors $\bp$ with $p_i\leq k_i$, and computes $\wfomc(\Psi,\bp)$ using \eqref{eq: DAG_Acyclic_essential} (with the change of variable from $m$ to $|\bk|-l$), and storing it in $A[\bp]$. The number of iterations is bounded by $n^{u\times|T|}\leq n^{ud^{u}}$. For each iteration, in order to compute $\wfomc(\Psi,\bp)$, it calls less than $n$ times the function $\overline{\fomc}(\Psi_{[m]},\bp)$ which computes $\fomc(\Psi_{[m]},\bp)$ using \eqref{eq: k_k'_m_essential}. Inside this function, the number of iterations of the for loop is bounded by $n^{2u\times |T|}\leq n^{2ud^{u}}$. Each iteration computes $N(\phi,\bs',\bs'')$ using \cref{eq:extensions_essential} and $\fomc(\Psi',\bs')$ using \cref{thm: beam}. Both can be done in polynomial time w.r.t.\ n, so this analysis proves that the entire algorithm has polynomial complexity w.r.t.\ $n$.

\begin{algorithm}[tb]
    \caption{FOMC with $EssentialDAG(R,d)$}
    \label{alg:algorithm_FO_DAG}
    \begin{algorithmic}[1]
    \State \textbf{Input}: $\Psi \coloneqq \forall xy. \phi(x,y) \land EssentialDAG(R, d), \bk$
    \State \textbf{Output}: $\fomc(\Psi,\bk)$
    \State
        \State $A[\mathbf{0}] \gets 1$ \Comment{$A$ has $u\times |T|$ indices, $\mathbf{0} = \langle 0,...,0 \rangle$}
        \For{$\mathbf{0} < \bp \leq \bk$} \Comment{Lexical order} 
        \State ${\!A[\bp] \gets \!\sum_{l=0}^{|\bp|-1}(-1)^{|\bp|-l+1}\binom{|\bp|}{l}{\overline{\fomc}}(\!\Psi_{[|\bp|-l]},\bp)}$ 
        \EndFor\\
        \textbf{return} $A[\bk]$ 
        \State
        \Function{$\overline{\fomc}$}{$\Psi_{[m]}$, $\bs$} \Comment{Equation \eqref{eq: k_k'_m_essential}}
        \State $S = 0$
        \For{$\bs' + \bs'' = \bs$ and $|\bs'| = m$}
        \State $S \gets S + N(\phi, \bs', \bs'')\fomc(\Psi',\bs') A[\bs'']$ 
        \EndFor
        \State \Return $S$
        \EndFunction
        \end{algorithmic}
\end{algorithm}

For simplicity of exposition we have considered FOMC. However, given a symmetric weight function, one could introduce weights of $1$-types and $2$-tables as done with \eqref{eq: w_i} and \eqref{eq: v_l}. Then, the only changes needed for transforming the previous propositions for weighted model counting are the following:
\begin{itemize}
    \item change FOMC to WFOMC inside the formulas;
    \item inside \cref{prop: extensions_k_essential}, define $c_{ij}$ as $\sum_{l\in [b]} c_{ijl}v_l$, and $d_{ij}$ as $\sum_{l\in [b]} d_{ijl}v_l$, with $v_l$ defined in \eqref{eq: v_l}.
\end{itemize}
Moreover, one can use the machinery discussed in the background section and introduced in \cite{broeck2013,kuzelka2020,AAAI_Sagar} for modularly reducing, in polynomial time, the WFOMC of any C$^2$ formula to a WFOMC problem with FO$^2$ formulas in Skolem normal form, for which \cref{alg:algorithm_FO_DAG} can be used. Recalling that the number of vectors $\bk$ summing to $n$ is polynomially bounded, we get the following final theorem:

\begin{theorem}
    For any sentence $\Phi$ in C$^2$, $$\wfomc(\Phi\land EssentialDAG(R,d),n)$$ can be computed in polynomial time with respect to the domain cardinality $n$.
\end{theorem}

\section{Counting Essential DAGs by Indegrees}
As a special case, let us derive a formula for counting essential DAGs with bounded indegrees from the ones for FOMC with the essential DAG axiom.
In order to do so, we consider the case in which $\phi = \top$ and there is only the binary relation $R$ and no other predicate. The $1$-type corresponding to $R(x,x)$ is always satisfied (since DAGs have no self-loops). This means that we can consider as if it were $u=1$, and the extended $1$-types were given just by the number $|\bt|$ --- but let us call it just $t$ --- of parents of a node. Hence, we can read the vector $\bk$ as a vector of length $d$ such that $k_t$ specifies how many nodes have indegree $t$, for each $t=1,\dots,d$. The same for $\bk'$ and $\bk''$, with the difference that they refer to a subset of the nodes. 
%
Because the $2$-tables are those listed in \cref{ex:2-tables}, we have that $c_{11} = d_{11} = 1$. Hence,
\eqref{eq:extensions_essential} can be simplified to
\begin{equation*}
    N(\top, \bk', \bk'') = \prod_{t=0}^d g(\bk'',t)^{k'_t}
 \end{equation*}
where
    $g(\bk'',t) = \binom{n-m}{t} - k''_{t-1}$

Now that we have analyzed \cref{prop: extensions_k_essential}, we can pass to \cref{prop:induction_essential}. In this setting, 
$\fomc(\Psi',\bk')=\binom{m}{\bk'}$, since the only way to satisfy $\Psi'$ is that no arrows are drawn, so we are left with counting the number of ways to distribute the extended $1$-types to the $m$ nodes. This leaves us with 
\begin{align*}
\begin{split}
    & \fomc(\Psi_{[m]},\bk) = \\ & = \sum_{\substack{\bk= \bk'+\bk''\\ |\bk'|=m}}\binom{m}{\bk'}\prod_{t=0}^d \left(\binom{n-m}{t} - k''_{t-1}\right)^{k'_t} \!\! \fomc(\Psi,\bk'')
\end{split}
\end{align*}
which can be substituted inside \eqref{eq: DAG_Acyclic_essential} to get the final recursive formula:
\begin{align*}
\begin{split}
    a_{\bk} = & \sum_{m=1}^{n}(-1)^{m + 1}\binom{n}{m} \cdot \\ & \cdot \sum_{\substack{\bk= \bk'+\bk''\\ |\bk'|=m}} \binom{m}{\bk'} \prod_{t=0}^d \left(\binom{n-m}{t} - k''_{t-1}\right)^{k'_t} a_{\bk''}
\end{split}
\end{align*}
where $a_\bk$ denotes the number of essential DAGs with cardinality vector of indegrees $\bk$.
\begin{table}
\centering
\begin{tabular}{c|ccccc}
n\textbackslash  d  & 2 & 3 & 4 & 5 \\
\midrule
3        & 4 & & &        \\
4        & 55 & 59 &   &     \\
5        & 1511 & 2341 & 2616  &      \\
6        & 68926 & 201666 & 292071   &  306117 \\
7        & 4724917 & 32268692 & 70992832  &  85672147 \\
8        & $4.5\times 10^8$ & $8.6 \times 10^9$ & $3.3\times 10^{10}$  &  $5.2\times 10^{10}$  \\
9        & $5.9\times 10^{10}$ & $3.6\times 10^{12}$ & $2.8\times 10^{13}$ & $6.5 \times 10^{13}$ \\
10       & $9.8\times 10^{12}$ & $2.2\times 10^{15}$ & $4.0\times 10^{16}$   & $1.5\times 10^{17}$ \\
11       & $2.0\times 10^{15}$ & $1.9\times 10^{18}$ & $9.0\times 10^{19}$   & $6.3\times 10^{20}$ \\
12       & $5.2\times 10^{17}$ & $2.2\times 10^{21}$ & $3.1\times 10^{23}$   & $4.4\times 10^{24}$ \\
\end{tabular}
\caption{Number of essential DAGs on $n$ nodes with indegree bounded by $d$. On the diagonal ($4, 59, 2616, 306117, \dots$), you can recognize the total number of essential DAGs with a given number of nodes, which could be computed using instead \cref{eq:counting_essential_DAGs}.}
\label{tab:n_d}
\end{table}
Summing $a_{\bk}$ over the vectors $\bk$ with $|\bk|=n$, we can count the number of essential DAGs on $n$ nodes with indegree bounded by $d$, with complexity exponential in $d$ but polynomial in $n$. Table \ref{tab:n_d} reports the results for $n\leq 12$ and $d\leq 5$. Moreover, the numbers $a_{\bk}$ allow for a very fine-grained counting. For example, by summing only over the vectors $\bk$ with $k_0=s$, one is constraining the essential DAG to have exactly $s$ sources. As another example, the number of edges of the essential DAG is a function of the vector $\bk$, so that any constraint on the number of edges can be 
encoded as a restriction on the vectors $\bk$ to be considered.


\section{Conclusion}
In this paper, we have introduced a novel approach to counting Markov Equivalence Classes (MECs) of DAGs under logical constraints. Specifically, our focus has been on MECs of size one, and we provided a polynomial-time algorithm for enumerating these essential DAGs with arbitrary logical constraints expressed in first-order logic with two variables and counting quantifiers (C$^2$). Overall, our work represent a step towards a deeper understanding of MECs. We integrate logical constraints and recent advancements in tractable first-order model counting with causal inference and the combinatorial analysis of MECs. Future work may extend our approach to MECs of arbitrary size, for example by taking inspiration from the way in which \cite{GILLISPIE2006} uses the counting of essential DAGs to count MECs of arbitrary size.

\bibliographystyle{kr}
\bibliography{references}

\end{document}